# An XMCD study of magnetism and valence state in iron-substituted strontium titanate

Astera S. Tang[†], Jonathan Pelliciari[†], Qi Song, Qian Song, Shuai Ning, John W. Freeland, Riccardo Comin, Caroline A. Ross

[†]These authors contributed equally

A.S. Tang, S. Ning, Prof. C.A. Ross

Department of Materials Science and Engineering, Massachusetts Institute of Technology, Cambridge MA 02139; astera@mit.edu, caross@mit.edu

J. Pelliciari, Q. Song, Q. Song, Prof. R. Comin

Department of Physics, Massachusetts Institute of Technology, Cambridge MA 02139; jpellici@mit.edu, rcomin@mit.edu

Q. Song

State Key Laboratory of Surface Physics and Department of Physics, Fudan University, Shanghai 200433, China

J.W. Freeland

Advanced Photon Source, Argonne National Laboratory, Argonne, Illinois, 60439, USA

**Abstract**

Room temperature ferromagnetism was characterized for thin films of $SrTi_{0.6}Fe_{0.4}O_{3-\delta}$ grown by pulsed laser deposition on $SrTiO_3$ and Si substrates under different oxygen pressures and after annealing under oxygen and vacuum conditions. X-ray magnetic circular dichroism demonstrated that the magnetization originated from $Fe^{2+}$ cations, whereas $Fe^{3+}$ and $Ti^{4+}$ did not contribute. Films with the highest magnetic moment (0.8 $\mu_B$ per Fe) had the highest measured $Fe^{2+}$:$Fe^{3+}$ ratio of 0.1 corresponding to the largest concentration of oxygen vacancies ($\delta = 0.19$).

Post-growth annealing treatments under oxidizing and reducing conditions demonstrated quenching and partial recovery of magnetism respectively, and a change in Fe valence states. The study elucidates the microscopic origin of magnetism in highly Fe-substituted $SrTi_{1-x}Fe_xO_{3-\delta}$ perovskite oxides and demonstrates that the magnetic moment, which correlates with the relative content of $Fe^{2+}$ and $Fe^{3+}$, can be controlled via the oxygen content, either during growth or by post-growth annealing.

## I. INTRODUCTION

Transition metal oxides are a versatile class of materials with collective charge and spin phenomena including ferro/antiferromagnetism, superconductivity, ferroelectricity, colossal magnetoresistance, and multiferroicity [1–10]. The perovskite structure offers great compositional flexibility owing to the ability of the cation sites to accommodate a range of ionic sizes and cation valencies, permitting fine control of electronic and magnetic properties [3,4,8]. A key aspect of the behavior is the role played by defects and vacancies, both of which can be intrinsic sources of conductivity, magnetism and other properties [11–14].

Fe-substituted $SrTiO_3$ (STF) is a material where room temperature ferromagnetism is introduced into a non-ferromagnetic $SrTiO_3$ (STO) host [15]. Structurally, Fe replaces Ti in the B site of the perovskite structure as schematically shown in Fig. 1(a). Two regimes may be distinguished: a dilute level of substitution in which nearest neighbor (Fe-O-Fe) cation configurations are rare, and a highly-substituted regime in which nearest neighbor configurations are common and exchange interactions between Fe cations become important. In bulk STF, the Fe ions typically exist with an $Fe^{3+}$ or $Fe^{4+}$ oxidation state [16,17]. However, thin film growth such as by pulsed laser deposition (PLD) under low oxygen pressures can lead to the presence of $Fe^{2+}$ [18–22] as observed by surface sensitive techniques, due to the kinetically-limited growth process which results in a high concentration of oxygen vacancies that are compensated by changes to the cation valence state [18]. Although x-ray photoelectron spectroscopy (XPS) studies of Fe-substituted STO have been conducted [15,20,21], there is no study of STF using both x-ray absorption (XAS) and x-ray magnetic circular dichroism (XMCD) which can provide a detailed microscopic mechanism for the emergence of magnetism with bulk sensitivity. XPS and XMCD studies on binary oxides such as magnetite [23] exist; however, these materials have qualitatively different structures compared to STF. A systematic investigation of STF with spectroscopic methods will play a key role in unveiling the mechanism responsible for magnetism.

STF and related materials such as Co-substituted STO have demonstrated room temperature ferromagnetism when grown under low oxygen partial pressures [12,15,20,21,24,25]. PLD-grown STF thin films on both Si and STO substrates exhibited room temperature ferromagnetism when deposited at pressures below 5 µTorr, in contrast to films grown at higher oxygen pressures [15,26]. The former study showed higher magnetization for films grown on Si as compared to those on $SrTiO_3$ for nominally identical deposition conditions, and found the presence of $Fe^{3+}$ but did not detect $Fe^{2+}$ through Mössbauer spectroscopy and XPS. Laser-irradiated undoped STO single crystals and thermally annealed Co-doped (La,Sr)$TiO_3$ also exhibited ferromagnetism, which was suggested to be related to oxygen vacancies [12,27]. In these materials, oxidizing treatments lowered the magnetic moment while reducing treatments increased it. Density functional theory calculations predict magnetism in $SrTiO_3$, Fe-substituted $SrTiO_3$, and Co-substituted $SrTiO_3$ that can arise from Ti vacancies, oxygen vacancies, or substituted transition metal cations [15,28,29].

Here, we report the evolution of Fe valence state and magnetic moment in highly Fe-substituted STO films which are grown at different oxygen pressures and on different substrates. By a combination of x-ray diffraction (XRD), transmission electron microscopy (TEM), vibrating sample magnetometry (VSM), XAS [30,31], and XMCD [31–33], we uncover the link between room temperature magnetization and the electronic configuration of Fe and Ti ions as a function of oxygen vacancy concentration. The magnetization increases with the $Fe^{2+}$:$Fe^{3+}$ ratio, and therefore with the oxygen vacancy concentration. We demonstrate that the magnetism can be quenched on annealing in an oxidizing environment and partly restored upon annealing in a reducing environment. The experimental evidence identifies the origin of magnetism in this class of materials as arising from the presence of $Fe^{2+}$ in an otherwise $Fe^{3+}$ system, which modifies the dominant antiferromagnetic $Fe^{3+}$ interactions. These results extend our understanding of the source of magnetism in Fe-substituted $SrTiO_3$, and are helpful in facilitating the design of oxide materials whose magnetic properties can be manipulated, e.g. by annealing or electrochemical means.

## II. EXPERIMENTAL DETAILS

STF films were deposited from a $SrTi_{0.6}Fe_{0.4}O_3$ target on single-crystal (100) Si substrates with a native oxide layer and on single-crystal (100) $SrTiO_3$ substrates using a Neocera pulsed laser deposition (PLD) system with a KrF laser (248 nm) and a fluence of 1.4 J/cm$^2$. A previous study [24] using the same target and nominally the same deposition conditions indicated that films have less Fe than the target composition, i.e. a 40% Fe target yielded films of composition $SrTi_{0.65}Fe_{0.35}O_{3-\delta}$. Measurement of a sample of this study grown on Si at 3 µTorr gave a ratio of Ti:Fe = 67:33. The substrate temperature was 650°C for all depositions, and the base and growth pressures varied between 1 µTorr-4 µTorr. Base pressure refers to the pressure in the chamber before the substrate heater was turned on, while growth pressure refers to the pressure in the chamber at the start of deposition after the sample reached the deposition temperature. All depositions were made on 10 mm × 10 mm × 0.5 mm thick substrates. Annealing treatments carried out in the PLD chamber consisted of: (1) oxidation at a substrate temperature of 650°C and pressure of 160 Torr of 100% $O_2$ for two hours, followed by (2) reduction under vacuum at a substrate temperature of 650°C and starting pressure of 1.5 µTorr for 45 minutes.

Results from nine selected samples are discussed in this article. Samples are referenced by their base and growth pressures as "$x$-$y$ µTorr" where $x$ is the base pressure in µTorr and $y$ is the growth pressure in µTorr. The films subjected to the annealing cycle are referred to as: "as-grown", "oxidized" representing the same sample after annealing in oxygen, and "reduced" representing the sample after oxidation followed by annealing in vacuum. The base pressures, growth pressures, thicknesses, and annealing treatment of the films are shown in Table 1. For the first six samples, each pair of films on Si and $SrTiO_3$ was grown during the same deposition process. All thicknesses were measured by profilometry.

XRD ω-2θ scans were performed using a Rigaku Smartlab Multipurpose Diffractometer with the Smartlab Guidance data collection program and an incident-beam Ge (022) double bounce monochromator. Reciprocal space mapping (RSM) scans were collected on a Bruker D8 High Resolution Diffractometer. For the films on STO, the unit cell volume was calculated on the basis that the in-plane lattice parameter matched that of the STO as shown by the RSM data (Supporting Information [40]) and in Refs. [15,21], and taking the out-of-plane lattice parameter from the highest intensity peak in the XRD data. For films on Si we assume a cubic unit cell, based on earlier work showing a *c/a* ratio of 0.990-0.998 for STF/Si [15]. TEM samples were prepared with a Helios Nanolab 600 Dual Beam Focused Ion Beam Milling System. TEM images were collected with a JEOL 2010 Advanced High Performance TEM at 200 kV and elemental mapping was performed with a JEOL 2010 FEG Analytical Electron Microscope at 250 kV. Magnetic hysteresis loops were measured using a Digital Measurement System 7035B vibrating sample magnetometer (VSM). Composition of 35%Fe and cell volume as calculated from XRD data were used for unit conversion between emu/cm$^3$ and $\mu_B$/Fe.

XAS is a spectroscopic technique sensitive to the valence state and local environment of the atoms [30,31], while XMCD is sensitive to magnetism, particularly the local magnetization [31–33]. One of the great advantages of XMCD is the possibility of detecting local magnetization on different atoms and uncovering the source of magnetism in complex multielement systems, and by combining this information with XAS, the valence and the associated magnetism can be studied with elemental sensitivity. We used beamline 4-ID-C at the Advanced Photon Source (APS) at Argonne National Laboratory and measured the XAS in total electron yield (TEY) (when electrical conductivity was sufficient) and total fluorescence yield (TFY) mode. All the XAS and XMCD data shown in the paper are TFY, which is more sensitive to the bulk of the sample rather than the surface. The samples were mounted vertically and the magnetic field was applied along the sample surface normal. A magnetic field of 3.36 kOe was applied by means of an octupole magnet and was oriented along the normal of the sample plane. The method for calculating the oxygen stoichiometry is described in Supporting Information [40].

**TABLE 1.** Base pressures, growth pressures, thicknesses, annealing conditions and unit cell volume from XRD and magnetic moment, fraction of $Fe^{2+}$, and $\delta$ for the $SrTi_{0.65}Fe_{0.35}O_{3-\delta}$ thin films on Si and on STO.

| Sample designation | Substrate | Base pressure, μTorr | Growth pressure, μTorr | Thickness, nm | Annealing conditions | Unit cell volume, Å³ | Magnetic moment (VSM), $\mu_B$ / Fe | Fraction $Fe^{2+}$ (XAS) | $\delta$ (XAS) |
|---|---|---|---|---|---|---|---|---|---|
| 1-1.2μTorr (Si) | Si | 1 | 1.2 | 129 | n/a | 60.85±0.87 | 0.81 | 8±3 % | 0.189±0.003 |
| 5-1.2μTorr (Si) | Si | 5 | 1.2 | 118 | n/a | 60.03±0.86 | 0.65 | 8±3 % | 0.189±0.003 |
| 2-2μTorr (Si) | Si | 2 | 2 | 124 | n/a | 60.15±0.86 | 0.43 | 8±3 % | 0.189±0.003 |
| 3-3μTorr (Si) | Si | 3 | 3 | 120 | n/a | 60.79±0.87 | 0.51 | 7±3 % | 0.187±0.003 |
| 4-4μTorr (Si) | Si | 4 | 4 | 173 | n/a | 60.10±0.86 | 0.28 | 5±3 % | 0.184±0.003 |
| 1-1.2μTorr (STO) | SrTiO₃ | 1 | 1.2 | 129 | n/a | 61.34±0.35 | 0.79 | 10±3 % | 0.192±0.003 |
| 5-1.2μTorr (STO) | SrTiO₃ | 5 | 1.2 | 118 | n/a | 60.45±0.19 | 0.074 | 0±3 % | 0.175±0.003 |
| 2-2μTorr (STO) | SrTiO₃ | 2 | 2 | 124 | n/a | 60.89±0.14 | 0.11 | 0±3 % | 0.175±0.003 |
| 1-1.5μTorr (STO): as-grown | SrTiO₃ | 1 | 1.5 | 102 | n/a | 61.45±0.15 | 0.84 | 10±3 % | 0.192±0.003 |
| oxidized | SrTiO₃ | 1 | 1.5 | 102 | 160 Torr/2 hr | 61.59±0.67 | 0.00 | 0±3 % | 0.175±0.003 |
| reduced | SrTiO₃ | 1 | 1.5 | 102 | 160 Torr/2 hr +1.5 μTorr/ 45 min | 61.24±0.18 | 0.66 | 6±3 % | 0.185±0.003 |

## III. RESULTS AND DISCUSSION

### A. Experimental Results

X-ray diffraction ω-2θ scans between 20°-80° are shown in Fig. 1(b)-(d). All films have a perovskite structure with no visible secondary phases. Polycrystalline perovskite films formed on the Si substrates in Fig. 1(b) and single-crystal films on the STO substrates in Fig. 1(c)-(d). The relative intensities of the peaks for STF on Si correspond to those of the polycrystalline reference for bulk cubic $SrTiO_3$, indicating no preferred texture. No metallic phases, iron oxides, titanium oxides, or strontium oxides were identified for the as-grown films or for the annealed films shown in Fig. 1(e). Multiple or asymmetrical (200) film peaks were seen in the ω-2θ scans which is indicative of a distribution of out-of-plane lattice parameters. Consistent with this result, RSM of films on STO showed that the film peak was broadened along $q_z$, but the film peak remained matched to the substrate peak along $q_x$ indicating a coherent interface with in-plane lattice matching to the substrate. TEM imaging of 1-1.2μTorr (STO) showed the presence of vertical planar defects in the single crystalline film; however, elemental mapping proved that the component elements including Fe were homogeneously distributed. Additional TEM images can be found in the Supporting Information [40].

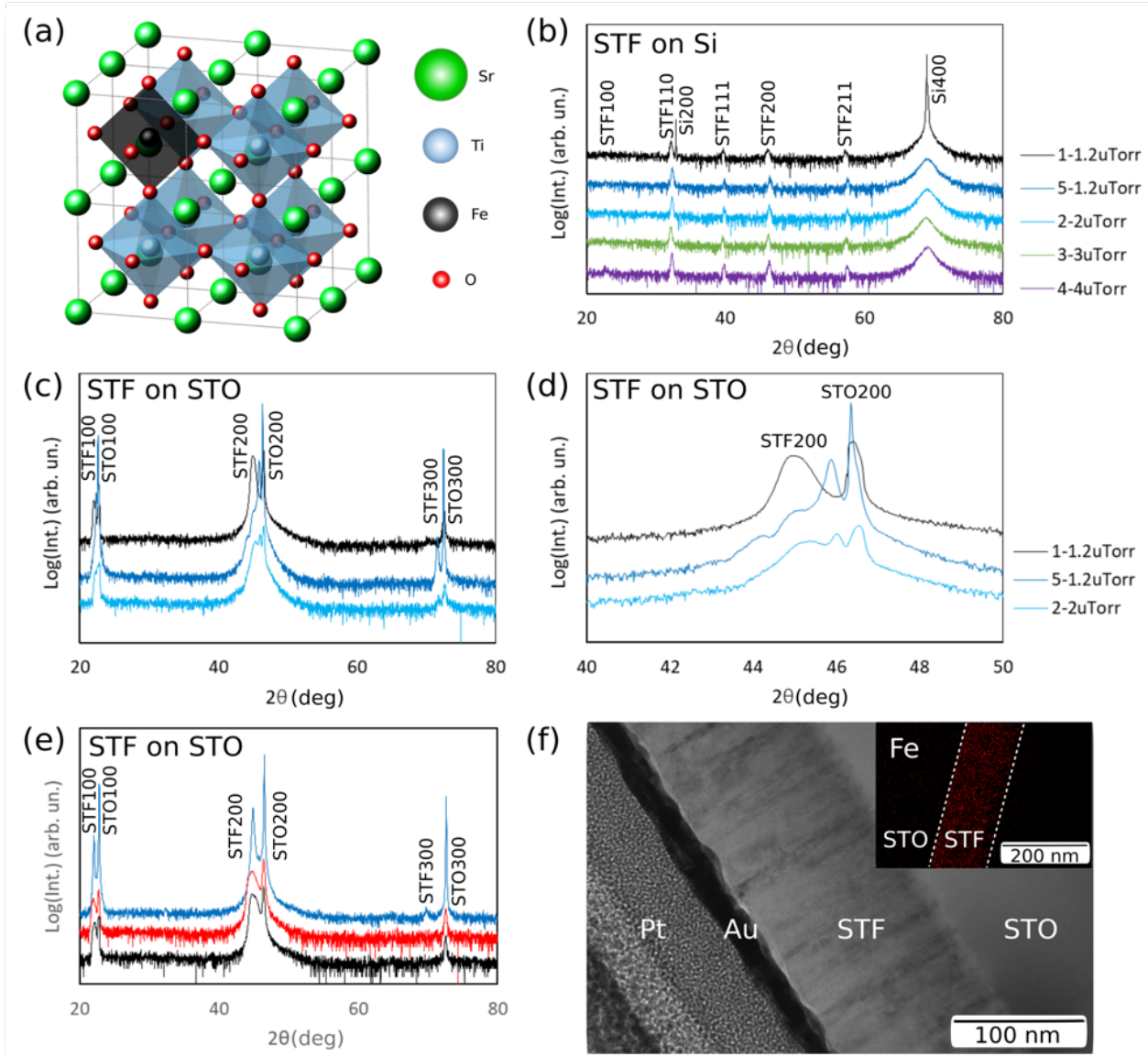

**FIG. 1.** (a) Model of Fe-substituted SrTiO$_3$ demonstrating cubic perovskite structure, oxygen octahedra surrounding the B sites, and cation substitution of Ti with Fe. (b-e) XRD ω-2θ scans for SrTi$_{60}$Fe$_{40}$O$_{3-\delta}$ samples on (b) Si and (c-e) SrTiO$_3$ substrates, subjected to different base and growth pressures (b-d) or annealing treatments (e). XRD ω-2θ scans between 40°<2θ<50° (d) are left of their corresponding wide-range scans (c). (f) TEM image of 1-1.2µTorr (STO) with Fe elemental map showing homogeneous distribution of Fe.

For oxygen-deficient STF, the presence of oxygen vacancies is compensated by lowering the oxidation state of the cations, leading to higher ionic radii and to chemical expansion of the unit cell compared to that of Sr(Ti,Fe)O$_3$ [15,21,35]. Unit cell volumes are shown in **Table 1**. As a comparison, bulk SrTi$_{0.65}$Fe$_{0.35}$O$_3$ without oxygen deficiency has an interpolated unit cell volume of 60.13Å$^3$. [15]

The 1-1.2µTorr (STO) film was grown at the lowest base pressure and deposition pressure and has the highest out-of-plane lattice parameter of the three films in Fig. 1(d), with $2\theta$=45.054° ($d_{hkl,002}$=2.011 Å, $c$=4.021±0.023 Å, error from peak full width at half maximum). The as-grown 1-1.5µTorr film had a lattice parameter of $c$ = 4.030±0.010 Å. These values are similar to that of the STF/STO sample with the highest magnetic moment reported in Ref. 15. Interestingly, the 5-1.2µTorr (STO) sample had a lower unit cell volume than the 1-1.2µTorr (STO) suggesting a role of base pressure in determining the film structure.

VSM hysteresis curves, XAS spectra, and XMCD hysteresis curves and spectra of the 1-1.2µTorr (STO) and 1-1.2µTorr (Si) films of Fig. 1(a,b) are compared in Fig. 2. It is apparent from VSM that both samples exhibit room temperature ferromagnetic behavior with an out-of-plane easy axis. The anisotropy in STF/STO has been attributed to magnetoelastic effects. [21,24,25,34] The two films exhibit a similar saturation magnetization ($M_s$) and remanence ($M_r$) despite their microstructural differences, with $M_s$ = 43 emu/cm$^3$ (0.81 µ$_B$/Fe) for STF/Si, $M_s$ = 42 emu/cm$^3$ (0.79 µ$_B$/Fe) for STF/STO, and $M_r$ = 33 emu/cm$^3$ (0.62 µ$_B$/Fe) for both. The coercivity $H_c$ for the film on Si, 2.3 kOe, is much higher than for the film on STO, 1.5 kOe. This may be a result of higher pinning at the grain boundaries of the polycrystalline film on Si compared to the single-crystalline film on STO.

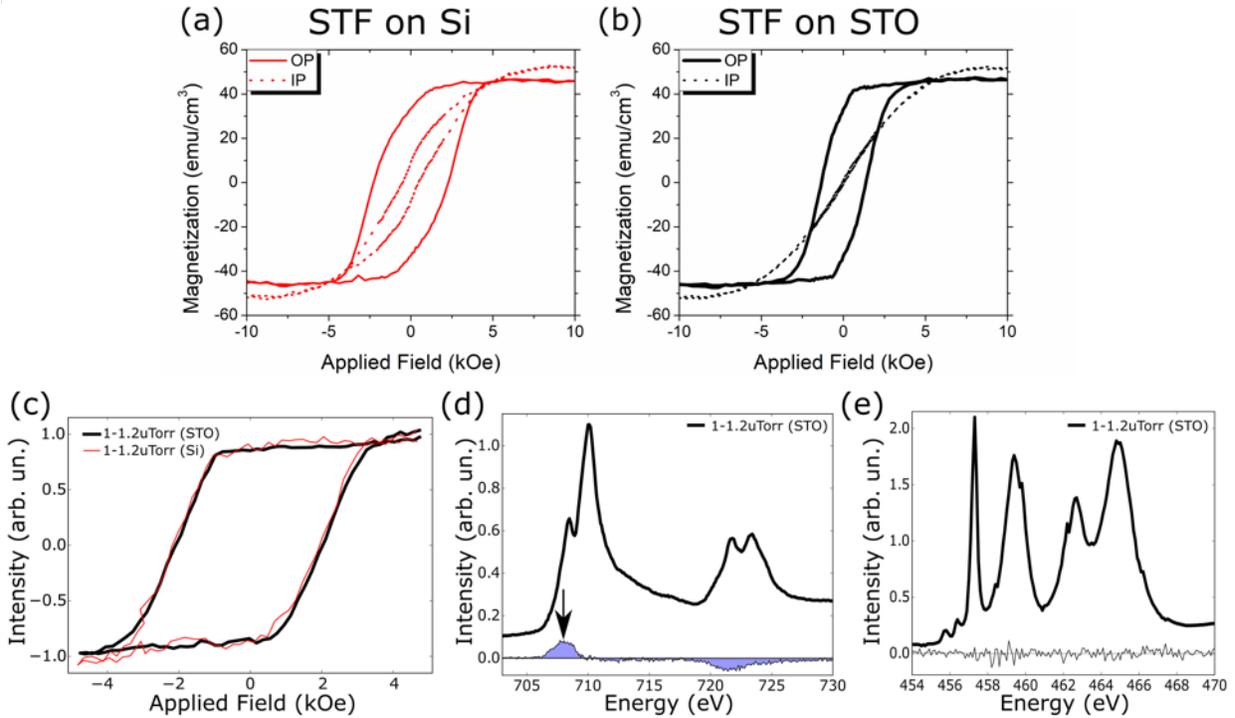

**FIG. 2.** (a,b) Out-of-plane (OP) and in-plane (IP) VSM hysteresis curves of $SrTi_{60}Fe_{40}O_{3-\delta}$ thin films grown at 1 µTorr base pressure, 1.2 µTorr growth pressure on (a) Si substrate and (b) $SrTiO_3$ substrate. (c) XMCD hysteresis curves for the same two films collected at the energy of the maximum Fe signal, shown with an arrow in (d). (d) Fe-L edge and (e) Ti-L edge XAS and XMCD spectra of $SrTi_{0.60}Fe_{0.40}O_{3-\delta}$ thin film grown at 1 µTorr base pressure, 1.2 µTorr growth pressure on $SrTiO_3$. All XAS and XMCD data shown here are TFY scans.

Figure 3(a-c) compares the out-of-plane VSM hysteresis loops for the series of films on both STO [Fig. 3(a)] and Si [Fig. 3(b)] as well as the XAS/XMCD spectra for the films on Si [Fig. 3(c)]. All the films grown on Si demonstrated room temperature ferromagnetism with out-of-plane easy axis, as exemplified by 1-1.2µTorr (Si) in Fig. 2(a), but the saturation magnetization and remanence decreased with increasing pressure during deposition. In comparison, the STF/STO films of Ref. 15 exhibited a maximum in magnetic moment and unit cell volume at a higher base pressure (3 – 4 µTorr) with a decrease in magnetic moment at higher pressures. Reference [15] also showed evidence of metallic Fe nanorods in films grown at low growth pressures, but metallic Fe was not observed in the present study. The differences may reflect the influence of beam focus, beam intensity, target condition or other deposition parameters on the film growth.

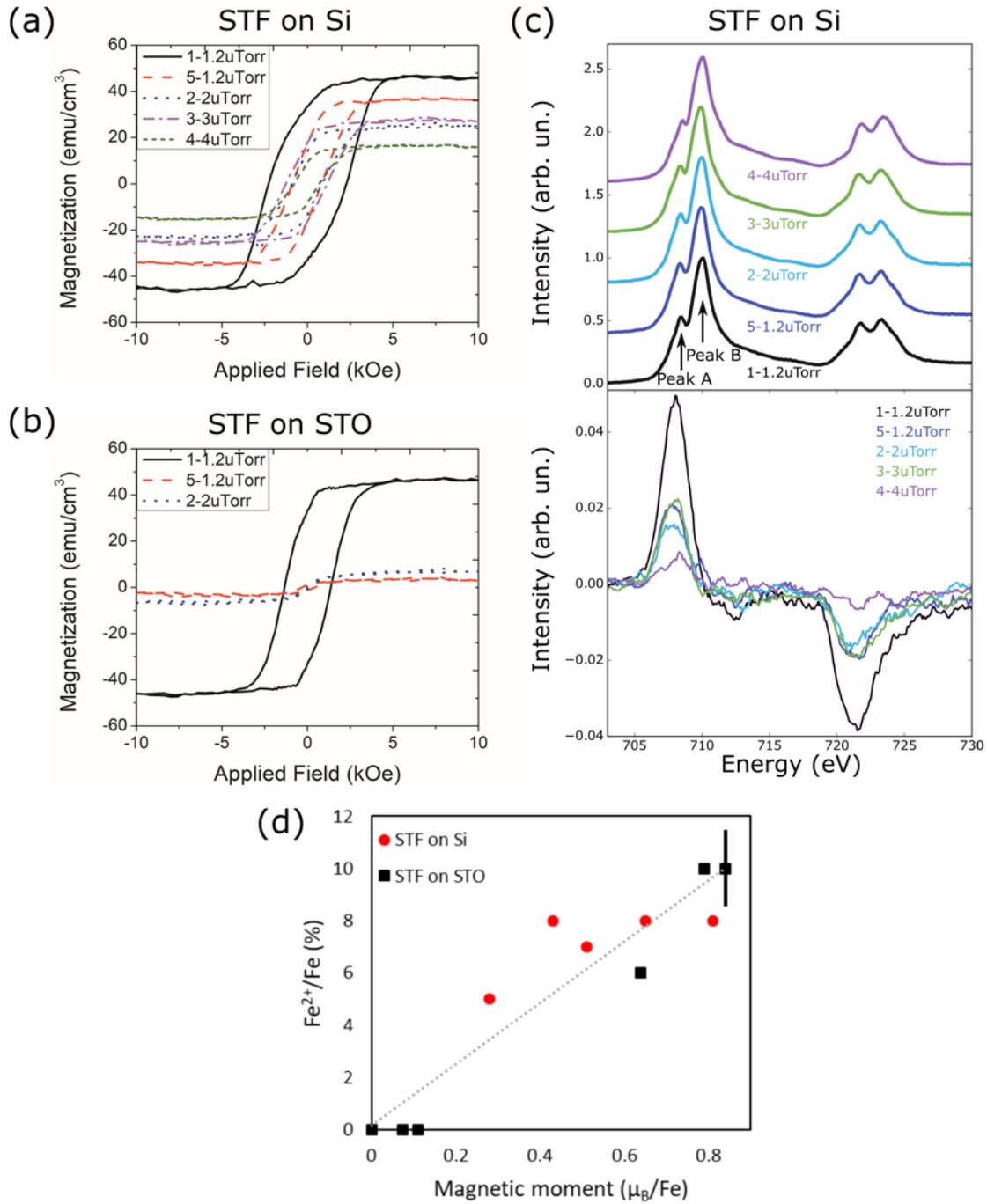

FIG. 3. (a,b) Out-of-plane VSM hysteresis curves of SrTi$_{60}$Fe$_{40}$O$_{3-\delta}$ thin films grown at different base and growth pressures on Si (a) and SrTiO$_3$ (b). (c) TFY Fe-L edge XAS and XMCD of SrTi$_{0.60}$Fe$_{0.40}$O$_{3-\delta}$ thin films grown on Si at different base and growth pressures. (d) The relation between the Fe valence states determined from XAS and the net magnetic moment measured by VSM. A representative error bar is shown.

Figure 2(c) displays XMCD hysteresis curves collected at the energy corresponding to the maximum Fe signal [marked with an arrow in Fig. 2(d)] for 1-1.2µTorr (STO) and 1-1.2µTorr (Si). The XMCD hysteresis curves for the two films are similar, unlike the VSM measurements. The XMCD measures a smaller area (1x1 mm$^2$) than the VSM (1x1 cm$^2$), so the difference in coercivity may be due to inhomogeneity in the film. VSM measurements on a smaller piece (5x5 mm$^2$) of the sample indicate some variation in coercivity within the sample (e.g. coercivity of 1700 Oe for the 5x5 mm$^2$ sample compared with 1400 Oe for the 1x1 cm$^2$ sample). The XMCD peak area of Fig. 3(c) scales with the VSM saturation magnetization, decreasing with increasing growth pressure (see Supporting Information [40]).

The lineshape of the Fe XMCD signal resembles the lineshape of Fe$^{2+}$ in FeTiO$_3$, however in our case the dilution and disorder of the Fe ions as well as the weak XMCD signal might hinder the detection of the multiplet structures detected in bulk FeTiO$_3$. [36] While similar XMCD signals are also observed in metallic Fe, an extensive search for Fe metal precipitates using x-ray diffraction, cross-sectional TEM, and x-ray photoelectron spectroscopy strongly hints at the absence of metallic Fe (see Supporting Information [40]). The lack of XMCD signal at the Ti-L edge in Fig. 2(e) is consistent with the magnetism residing at the Fe, specifically Fe$^{2+}$, rather than the Ti site as found in oxygen-deficient STO [27,28].

Having established the trend in the magnetic moment of STF vs. deposition pressure, we turn to XAS to examine the Fe valence states that give rise to the magnetism. The XAS spectrum, Fig. 2(d), reveals that the film exhibits a mixture of Fe$^{2+}$ and Fe$^{3+}$ (for an extensive comparison of the valence state of Fe see Supporting Information [40]). No Fe$^{4+}$ was detected, unlike previous studies [15,21,24,26] on ferromagnetic STF, and similarly there was no signal from metallic Fe. The Ti is present in a 4+ oxidation state [Fig. 2(e)]. The observation of a mix of $Fe^{2+}$ and $Fe^{3+}$ is consistent with earlier studies of mixed cation valencies in ferromagnetic STF films [15,20,21,24] and with the presence of oxygen vacancies.

To quantify the Fe valence states, we performed a principal component analysis of the $Fe^{2+}$ and $Fe^{3+}$ spectral fingerprints. We extract a phenomenological parameter, the ratio between the XAS features labeled peak A and peak B in Fig. S3, which is a proxy for the $Fe^{2+}$:$Fe^{3+}$ ratio, as detailed in Supporting Information Fig. S4 [40]. From the peak ratio of the measured XAS spectra, we estimate the $Fe^{2+}$:$Fe^{3+}$ ratio of each film. The results are summarized in Table 1.

The XAS data reveal a striking correlation between the Fe valence states and the net magnetic moment as measured by VSM, Fig. 3(d). Films grown under different base and deposition pressures showed a general trend of decreasing $Fe^{2+}$ fraction with decreasing magnetic moment. The most magnetic samples, 1-1.2µTorr (STO), 1-1.5µTorr (STO) and 1-1.2µTorr (Si), consisted of ~ 90 % $Fe^{3+}$ and 8-10 % (±3%) $Fe^{2+}$. In contrast, the oxidized 1-1.5µTorr (STO) and the 2-2µTorr (STO) films had little or no magnetic moment and no measurable $Fe^{2+}$.

Furthermore, the Fe valence state analysis can be used to infer the fraction of oxygen vacancies $\delta$ based on the composition $SrTi_{0.65}Fe_{0.35}O_{3-\delta}$. We assume that the ions are present as $Fe^{2+}$, $Fe^{3+}$, $Ti^{4+}$, and $O^{2-}$ (not considering fractional oxidation states); all the vacancies are doubly ionized; and that there is an insignificant concentration of free electrons and other defects. This yields $\delta = 0.192$ for the 1-1.2µTorr (STO) and 1-1.5µTorr (STO) samples, $\delta = 0.189$ for the 1-1.2µTorr (Si) sample, and $\delta = 0.175$ in films with no net magnetization (Table 1). The higher fraction of $Fe^{3+}$ at higher deposition pressures is consistent with a reduction in the concentration of oxygen vacancies [16,17].

In Fig. 4 we focus on the effect of annealing treatments, in particular the reversibility of the process. Figure 4(a) reports the ω-2θ scans between 40°-50°, Fig. 4(b,c) highlight the out-of-plane and in-plane VSM loops, and Fig. 4(d,e) depict XAS and XMCD data. As with the samples grown at differing pressures, multiple or asymmetrical (200) film peaks are indicative of a distribution of out-of-plane lattice parameters. RSM scans likewise show the film peak spread along $q_z$, while it it is aligned with the substrate peak along $q_x$ indicating a coherent interface. In prior work on perovskite films, the presence of multiple peaks was attributed to partial strain relaxation [34]. The oxidized film has a broader peak and a higher out-of-plane lattice parameter than the reduced film.

The sample was initially magnetic with a moment of 0.84 $\mu_B$/Fe, but annealing in 160 mTorr oxygen lowered the magnetization to zero, while annealing in vacuum (1.5 µTorr) partially restored the magnetic moment. These magnetometry observations are corroborated by the XMCD data. The loop shapes also changed irreversibly after the oxidation/reduction cycle with the out-of-plane loop having lower squareness and the in-plane loop showing a small hysteresis compared to the as-grown sample. The data demonstrate incomplete reversibility of the changes resulting from oxidation, similar to the irreversibility observed in thermally annealed Co-doped $(La,Sr)TiO_3$. [12] Consistent with the trends from as-grown films, the $Fe^{2+}$ content was lowered on oxidation and increased when the film was vacuum annealed. The annealing treatments conducted in this study therefore demonstrate the efficacy of post-growth processing as a technique to control ferromagnetism in highly-substituted STF.

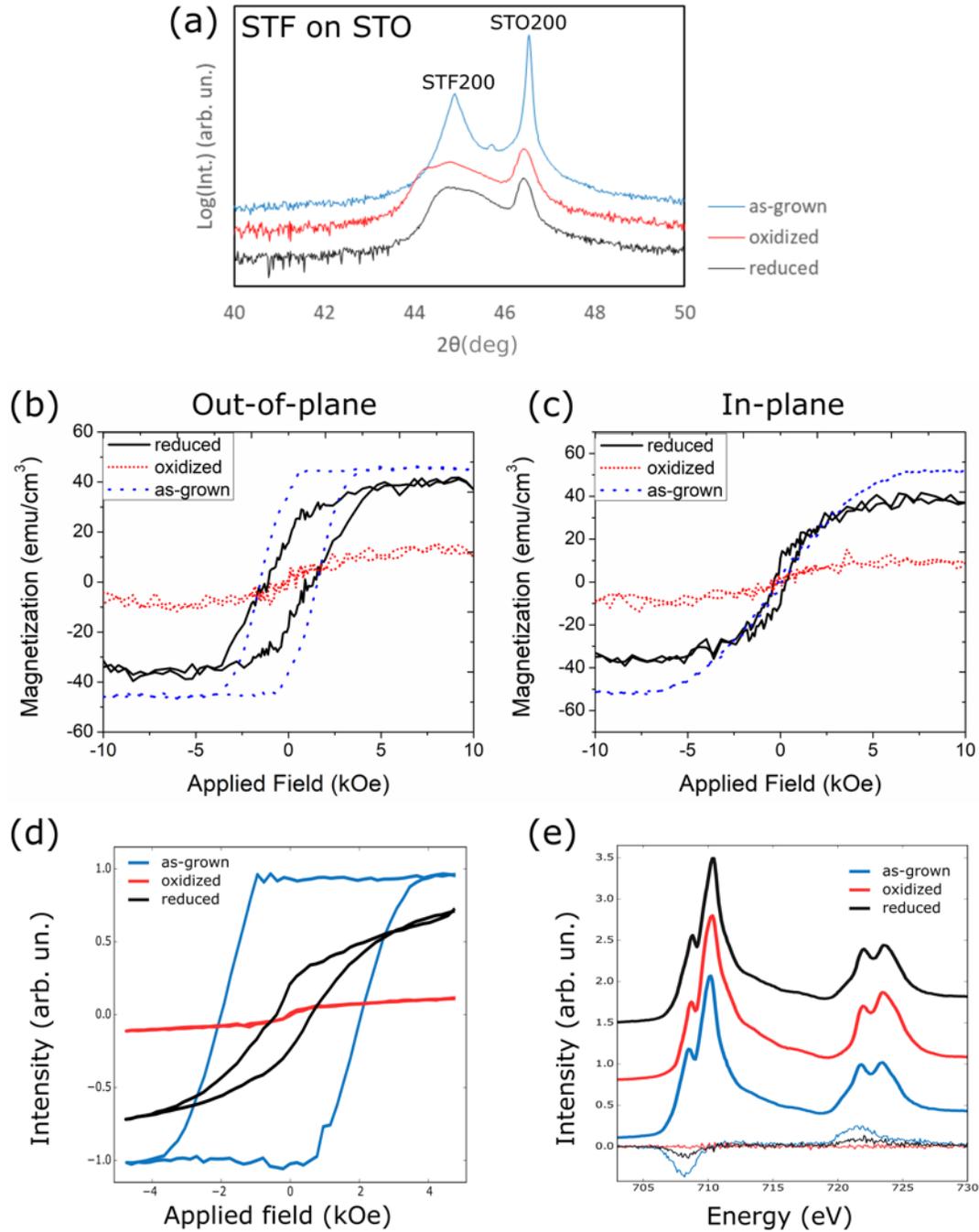

FIG. 4. (a) XRD ω-2θ scans between 40°<$2\theta$<50°, (b) out-of-plane and (c) in-plane VSM hysteresis curves for the SrTi$_{65}$Fe$_{35}$O$_{3-\delta}$ thin film subjected to various annealing treatments. (d) XMCD hysteresis curves collected on Fe signal and (e) XAS and XMCD for the same film subjected to various annealing treatments.

## B. Mechanism for Room Temperature Ferromagnetism

The previous analysis has shown that regardless of substrate, room temperature ferromagnetism in these STF films is correlated with the presence of $Fe^{2+}$, whose concentration can be controlled either during or after growth via manipulation of the oxygen content. Prior studies have suggested the importance of mixed cation valencies in room temperature ferromagnetism [14,15,20,24,26], and our results support this model in which the magnetic samples contain a mixture of $Fe^{2+}$ and $Fe^{3+}$. While we are unable to exclude the possibility of the oxygen vacancy itself also contributing to the magnetism [12,27,28,37], we note that even the non-magnetic samples include oxygen vacancies, suggesting that this contribution is minor. This conclusion is consistent with a study on La- and Ce-doped STF in which net magnetization increased with La-doping, which lowered the Fe valence state from 3+ to 2+ without nominally changing oxygen vacancy content [20]. Although the sensitivity of the magnetic moment to the $Fe^{2+}$ content is large, the actual change in $\delta$ required to balance the $Fe^{3+} \rightarrow Fe^{2+}$ valence state change is small because of the low fraction of $Fe^{2+}$ and the presence of only 35% of Fe on the B-sites. This could also explain why the unit cell volume was similar across the set of samples.

Magnetism in oxides has been explained from superexchange interactions as semi-empirically formulated in the Goodenough-Kanamori-Anderson rules [38,39] or from carrier-mediated mechanisms [9] originally developed to describe dilute magnetic semiconductors, such as double exchange or Ruderman-Kittel-Kasuya-Yosida (RKKY) coupling. Well-known carrier-mediated mechanisms were ruled unlikely in earlier studies on STF where it was noted that ferromagnetism persisted in highly insulating films [14,26]. Consequently, we focus on mechanisms that are more likely to dominate magnetism at the levels of B-site substitution present in this study.

The decrease of the magnetization with increasing growth or annealing pressure and the consequent lowering of oxygen vacancy concentration can be explained if one considers that neighboring $Fe^{3+}$ ions at the B-site of perovskites align antiferromagnetically as suggested by the Goodenough-Kanamori-Anderson rules [38,39] and as reported in DFT calculations on STF. [15] Since $Fe^{4+}$ was not detected in the XAS data, we consider mechanisms involving only $Fe^{2+}$ and $Fe^{3+}$. Typically, octahedrally-coordinated $Fe^{3+}$ in perovskites is present in a high spin state ($3d^5$, $S = 5/2$), resulting in a magnetic moment of 5 $\mu_B$/Fe. Without $Fe^{4+}$ or $Fe^{2+}$, compensation of the antiferromagnetically arranged spins occurs, obliterating any XMCD or VSM signatures of remanent magnetization. We assume that the $Fe^{2+}$ is also high spin. The presence of low spin $Fe^{2+}$ ($3d^6$, 0 $\mu_B$) is inconsistent with the XMCD result that the magnetism resides on the $Fe^{2+}$ ions rather than on the $Fe^{3+}$. Furthermore, it is unusual to find low spin $Fe^{2+}$ in the octahedral site [21,38], and low spin $Fe^{2+}$ (like high spin $Fe^{3+}$) is a non-magnetoelastic ion [21] which cannot account for the out-of-plane anisotropy observed in the magnetic films.

When $Fe^{2+}$ replaces $Fe^{3+}$, we can calculate the fraction of $Fe^{3+}$ that must be reduced to $Fe^{2+}$ to give the experimentally observed magnetization by considering the ratio of the magnetic moment measured via VSM to the theoretical moment of high spin $Fe^{2+}$ and $Fe^{3+}$ ions. The measured moment for 1-1.2µTorr (STO) according to VSM is 0.8 $\mu_B$/Fe. A high-spin $Fe^{2+}$ ion ($3d^6$, $S = 2$) is magnetoelastic [21] and has a magnetic moment of 4 $\mu_B$/Fe. In this case, magnetism could originate from $Fe^{2+}$ ions embedded in an antiferromagnetic $Fe^{3+}$ sublattice resulting in a net magnetization of around 0.4 $\mu_B$ (at 10% $Fe^{2+}$). While the resulting magnetization is not high enough to explain the experimentally observed magnetic moment given the concentration of $Fe^{2+}$, this discrepancy is resolved if the presence of an adjacent oxygen vacancy enhances the Fe moment [15]. The $Fe^{2+}$ content may also be higher than the values reported in Table 2 as a result of the linear calibration being affected by interference effects in the XAS signal. These effects are difficult to account for, and hence are not usually considered in these calculations.

Among the different superexchange interactions, it is possible to attain ferromagnetic alignment through exchange interactions with non-180° bond angles or non-/semi- covalent bond lengths [38,39], which is facilitated by the presence of a neighboring oxygen vacancy that distorts the octahedral symmetry surrounding the Fe. This can change the interaction between $Fe^{2+}$ and $Fe^{3+}$ to be ferromagnetic, or the distortion could change the crystal field splitting of neighboring Fe and result in some low spin ions.

Although the quantity of Fe and oxygen vacancies is high, there is no evidence that they will necessarily form an ordered sublattice. With a random or disordered distribution of Fe ions and oxygen vacancies in the STF lattice, we can expect a range of interactions among $Fe^{2+}$ and $Fe^{3+}$ ions through $O^{2-}$ or possibly $V_O^{\bullet\bullet}$ that will result in competing antiferromagnetic and ferromagnetic interactions, as well as differing enhancements to the Fe magnetic moment. The magnetism likely stems from some mix of the aforementioned mechanisms, which all rely on the mixed-valence state identified here and in prior studies [14,15,20,24,26] as crucial to inducing room temperature ferromagnetism.

## IV. CONCLUSION

To summarize, we have described the evolution of room temperature ferromagnetism in PLD-grown single-crystalline and polycrystalline $SrTi_{0.65}Fe_{0.35}O_{3-\delta}$ thin films as a function of deposition and annealing pressure. The key result is that the net magnetic moment, measured by magnetometry and by XMCD, showed a strong correlation with the $Fe^{2+}$ content as measured by XAS, and decreased with increasing growth pressure and with oxygen annealing. A principal component analysis is used to establish an empirical relationship between the ratio of the two XAS peak areas and the $Fe^{2+}:Fe^{3+}$ ratio. We were able to quantify the fraction of $Fe^{2+}$ through XAS and relate it to the observed VSM results, and subsequently derive the concentration of oxygen vacancies required for charge balance. We further observed the quenching and partial recovery of magnetization in a $SrTi_{0.65}Fe_{0.35}O_{3-\delta}$ thin film upon a cycle of oxidizing and reducing anneals, confirming that annealing can be used to actively change the magnetic moment and paving the way for studies utilizing electrochemical manipulation of the oxygen content.

The major source of magnetism in substituted STF appears to be the mixed Fe valence state consisting of $Fe^{2+}$ and $Fe^{3+}$, and putative mechanisms for the emergence of ferromagnetism are discussed. The presence of $Fe^{2+}$ disrupts the antiferromagnetic interactions between $Fe^{3+}$, facilitating a number of ferro- and ferrimagnetic interactions, and the presence of oxygen vacancies may enhance the resulting magnetic moment. The sensitivity of the magnetism to the amount of $Fe^{2+}$ and to small changes in base and growth pressures is surprisingly high. Our study underscores the importance of oxygen vacancies in highly-substituted transition metal oxides, and their powerful role in determining magnetization. The present study also motivates other methods to increase the $Fe^{2+}:Fe^{3+}$ ratio such as co-substitution of A-site or B-site cations with higher valence states than that of $Sr^{2+}$. Post-growth techniques can provide an additional pathway to engineer materials with desirable properties and also facilitate new applications. XMCD at high magnetic field or magnetic scattering studies could clarify the nature of the exchange interactions and the spin state of the ion species in the class of heavily magnetically-substituted perovskites.

**ACKNOWLEDGEMENT**

The authors acknowledge the support of MRL (formerly CMSE), an NSF MRSEC with award number DMR 1419807, and thank Dr. Yong Zhang for assistance with TEM. This work made use of the Shared Experimental Facilities supported by DMR 1419807. J. P. acknowledges financial support by the Swiss National Science Foundation Early Postdoc. Mobility and Mobility fellowship project numbers P2FRP2_171824 and P400P2_180744. Q. S. acknowledges financial support by the China Scholarship Council, number 201706100055.

**References**

[1] H. Y. Hwang, Y. Iwasa, M. Kawasaki, B. Keimer, N. Nagaosa, and Y. Tokura, Nat. Mater. **11**, 103 (2012).

[2] M. Kubicek, A. H. Bork, and J. L. M. Rupp, J. Mater. Chem. A **5**, 11983 (2017).

[3] G. Catalan, Phase Transitions **81**, 729 (2008).

[4] S. Catalano, M. Gibert, J. Fowlie, J. Íñiguez, J.-M. Triscone, and J. Kreisel, Reports Prog. Phys. **81**, 046501 (2018).

[5] H. Suzuki, H. Bando, Y. Ootuka, I. H. Inoue, T. Yamamoto, K. Takahashi, and Y. Nishihara, J. Phys. Soc. Japan **65**, 1529 (1996).

[6] L. W. Martin and A. M. Rappe, Nat. Rev. Mater. **2**, (2016).

[7] W. Prellier, M. P. Singh, and P. Murugavel, J. Phys. Condens. Matter **17**, R803 (2005).

[8] P. Zubko, S. Gariglio, M. Gabay, P. Ghosez, and J.-M. Triscone, Annu. Rev. Condens. Matter Phys. **2**, 141 (2011).

[9] N. Izyumskaya, Y. Alivov, and H. Morkoç, Crit. Rev. Solid State Mater. Sci. **34**, 89 (2009).

[10] J. M. Rondinelli and N. A. Spaldin, Adv. Mater. **23**, 3363 (2011).

[11] R. Moos and K. H. Härdtl, J. Am. Ceram. Soc. **80**, 2549 (1997).

[12] T. Fix, M. Liberati, H. Aubriet, S.-L. Sahonta, R. Bali, C. Becker, D. Ruch, J. L. MacManus-Driscoll, E. Arenholz, and M. G. Blamire, New J. Phys. **11**, 1 (2009).

[13] E. Enriquez, A. Chen, Z. Harrell, P. Dowden, N. Koskelo, J. Roback, M. Janoschek, C. Chen, and Q. Jia, Sci. Rep. **7**, 1 (2017).

[14] M. Egilmez, G. W. Leung, A. M. H. R. Hakimi, and M. G. Blamire, J. Appl. Phys. **108**, (2010).

[15] T. Goto, D. H. Kim, X. Sun, M. C. Onbasli, J. M. Florez, S. P. Ong, P. Vargas, K. Ackland, P. Stamenov, N. M. Aimon, M. Inoue, H. L. Tuller, G. F. Dionne, J. M. D. Coey, and C. A. Ross, Phys. Rev. Appl. **7**, 024006 (2017).

[16] A. Rothschild, W. Menesklou, H. L. Tuller, and E. Ivers-Tiffée, Chem. Mater. **18**, 3651

(2006).

[17] J. J. Kim, M. Kuhn, S. R. Bishop, and H. L. Tuller, Solid State Ionics **230**, 2 (2013).

[18] A. Koehl, D. Kajewski, J. Kubacki, C. Lenser, R. Dittmann, P. Meuffels, K. Szot, R. Waser, and J. Szade, Phys. Chem. Chem. Phys. **15**, 8311 (2013).

[19] C. Lenser, A. Kuzmin, J. Purans, A. Kalinko, R. Waser, and R. Dittmann, J. Appl. Phys. **111**, 109 (2012).

[20] P. Jiang, L. Bi, X. Sun, D. H. Kim, D. Jiang, G. Wu, G. F. Dionne, and C. A. Ross, Inorg. Chem. **51**, 13245 (2012).

[21] D. H. Kim, L. Bi, P. Jiang, G. F. Dionne, and C. A. Ross, Phys. Rev. B **84**, 014416 (2011).

[22] J. Szade, K. Szot, M. Kulpa, J. Kubacki, C. Lenser, R. Dittmann, and R. Waser, Phase Transitions **84**, 489 (2011).

[23] Y. Lu, J. S. Claydon, E. Ahmad, Y. Xu, S. M. Thompson, K. Wilson, and G. Van Der Laan, Ieee Trans. Magn. **41**, 2808 (2005).

[24] D. H. Kim, N. M. Aimon, L. Bi, J. M. Florez, G. F. Dionne, and C. A. Ross, J. Phys. Condens. Matter **25**, 026002 (2013).

[25] L. Bi, H.-S. Kim, G. F. Dionne, and C. A. Ross, New J. Phys. **12**, 043044 (2010).

[26] H.-S. Kim, L. Bi, G. F. Dionne, and C. A. Ross, Appl. Phys. Lett. **93**, 092506 (2008).

[27] S. S. Rao, Y. F. Lee, J. T. Prater, A. I. Smirnov, and J. Narayan, Appl. Phys. Lett. **105**, (2014).

[28] Y. Zhang, J. Hu, E. Cao, L. Sun, and H. Qin, J. Magn. Magn. Mater. **324**, 1770 (2012).

[29] J. M. Florez, S. P. Ong, M. C. Onbaşli, G. F. Dionne, P. Vargas, G. Ceder, and C. A. Ross, Appl. Phys. Lett. **100**, 252904 (2012).

[30] F. De Groot, Coord. Chem. Rev. **249**, 31 (2005).

[31] F. De Groot and A. Kotani, *Core Level Spectroscopy of Solids*, 1st ed. (CRC Press, Boca Raton, 2008).

[32] T. Funk, A. Deb, S. J. George, H. Wang, and S. P. Cramer, Coord. Chem. Rev. **249**, 3 (2005).

[33] G. van der Laan and A. I. Figueroa, Coord. Chem. Rev. **277**, 95 (2014).

[34] P. Jiang, L. Bi, D. H. Kim, G. F. Dionne, and C. A. Ross, Appl. Phys. Lett. **98**, Appl. Phys. Lett. 98 231909 (2011) (2011).

[35] N. H. Perry, J. J. Kim, S. R. Bishop, and H. L. Tuller, J. Mater. Chem. A **3**, 3602 (2015).

[36] T. Fujii, M. Yamashita, S. Fujimori, Y. Saitoh, T. Nakamura, K. Kobayashi, and J. Takada, J. Magn. Magn. Mater. **310**, e555 (2007).

[37] T. Tietze, M. Gacic, G. Schütz, G. Jakob, S. Brück, and E. Goering, New J. Phys. **10**, (2008).

[38] J. Kanamori, J. Phys. Chem. Solids **10**, 87 (1959).

[39] J. B. Goodenough, Phys. Rev. **100**, 564 (1955).

[40] See Supplementary Material at [URL will be inserted by publisher] for additional details and data regarding the method for calculating oxygen stoichiometry, XAS and XMCD, TEM and elemental mapping, Raman spectra, and RSM

**Figures**

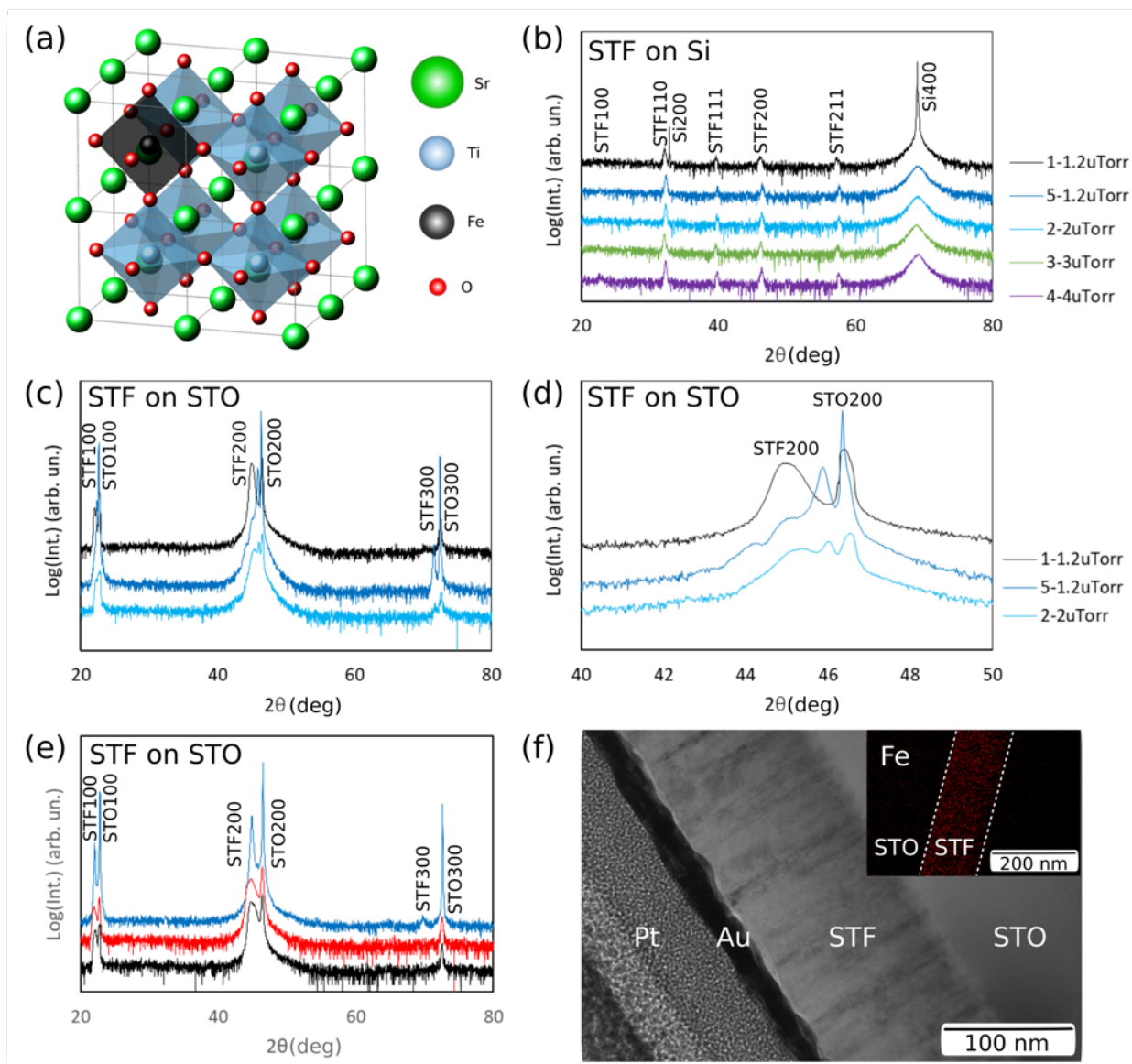

FIG. 1. (a) Model of Fe-substituted SrTiO$_3$ demonstrating cubic perovskite structure, oxygen octahedra surrounding the B sites, and cation substitution of Ti with Fe. (b-e) XRD ω-2θ scans for SrTi$_{60}$Fe$_{40}$O$_{3-\delta}$ samples on (b) Si and (c-e) SrTiO$_3$ substrates, subjected to different base and growth pressures (b-d) or annealing treatments (e). XRD ω-2θ scans between 40°<2θ<50° (d) are left of their corresponding wide-range scans (c). (f) TEM image of 1-1.2μTorr (STO) with Fe elemental map showing homogeneous distribution of Fe.

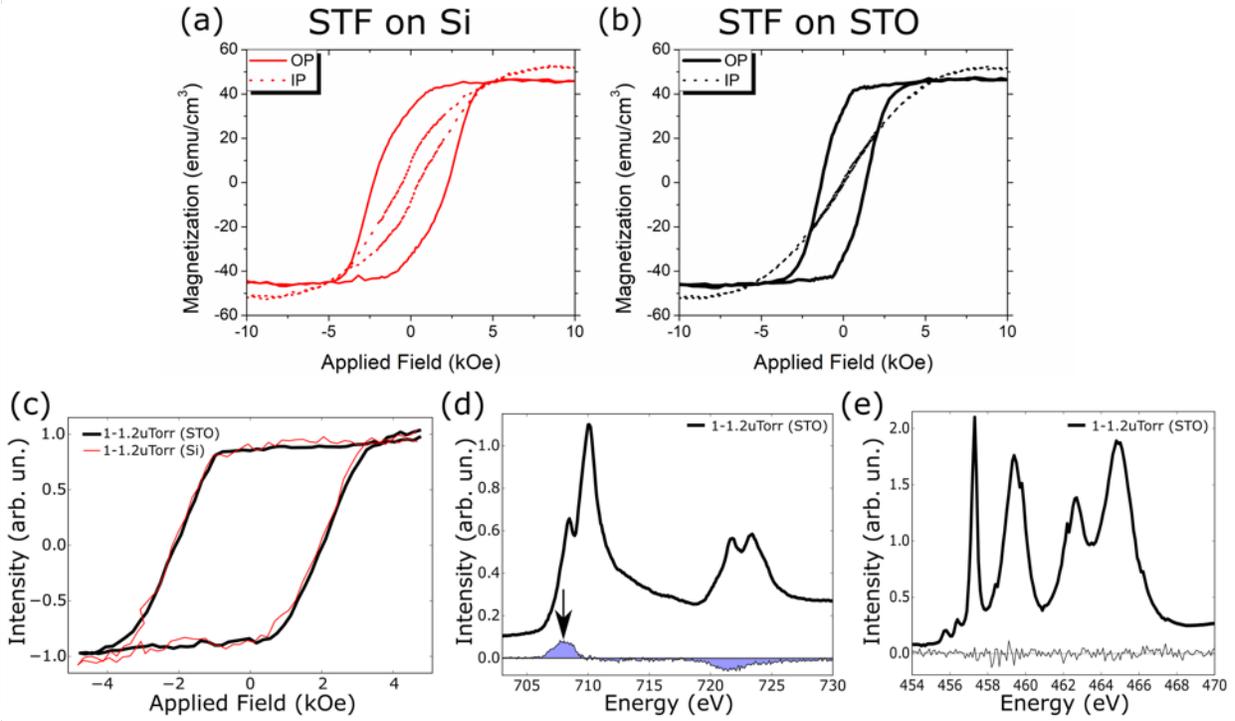

FIG. 2. (a,b) Out-of-plane (OP) and in-plane (IP) VSM hysteresis curves of $SrTi_{60}Fe_{40}O_{3-\delta}$ thin films grown at 1 µTorr base pressure, 1.2 µTorr growth pressure on (a) Si substrate and (b) $SrTiO_3$ substrate. (c) XMCD hysteresis curves for the same two films collected at the energy of the maximum Fe signal, shown with an arrow in (d). (d) Fe-L edge and (e) Ti-L edge XAS and XMCD spectra of $SrTi_{0.60}Fe_{0.40}O_{3-\delta}$ thin film grown at 1 µTorr base pressure, 1.2 µTorr growth pressure on $SrTiO_3$. All XAS and XMCD data shown here are TFY scans.

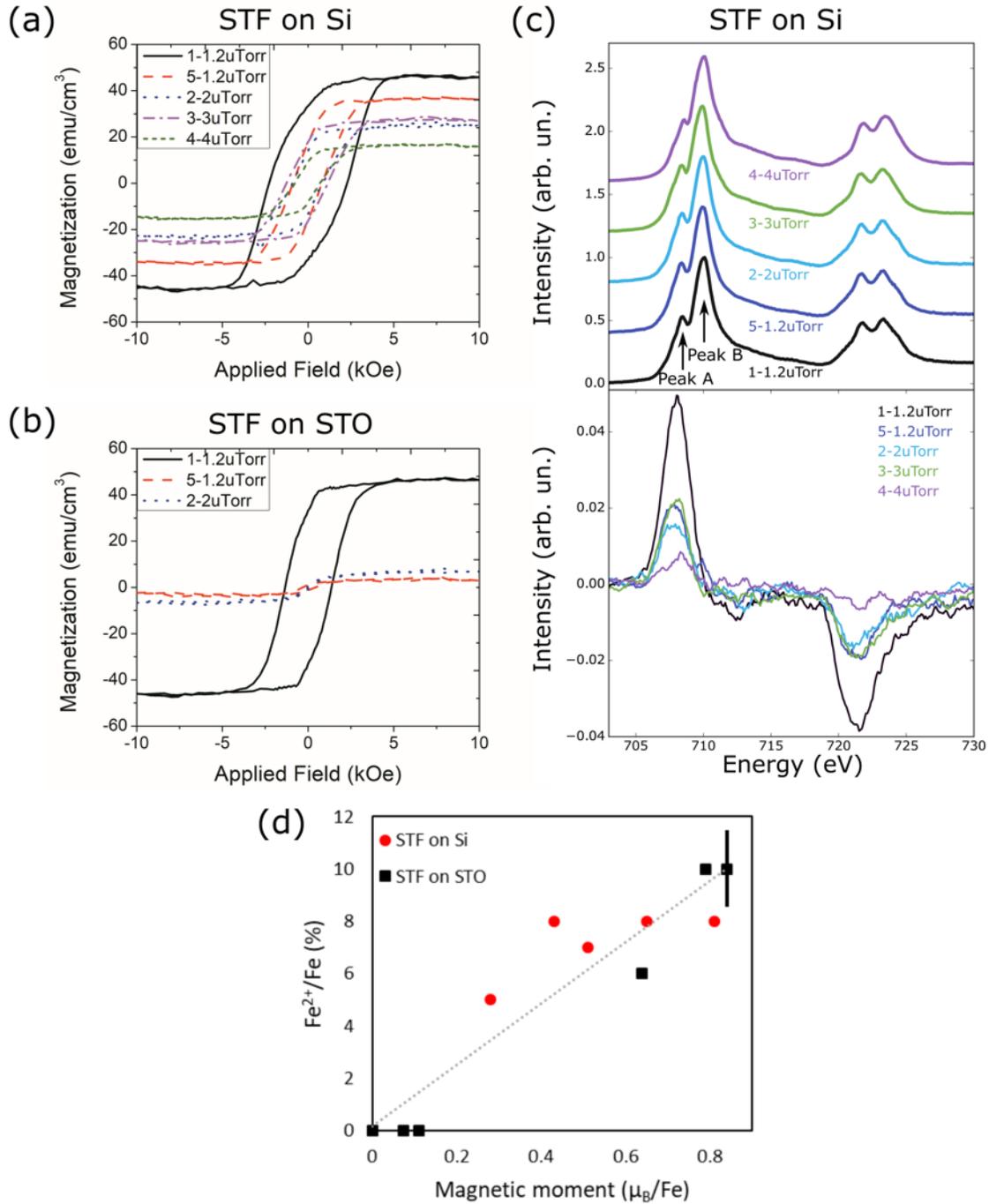

FIG. 3. (a,b) Out-of-plane VSM hysteresis curves of $SrTi_{60}Fe_{40}O_{3-\delta}$ thin films grown at different base and growth pressures on Si (a) and $SrTiO_3$ (b). (c) TFY Fe-L edge XAS and XMCD of $SrTi_{0.60}Fe_{0.40}O_{3-\delta}$ thin films grown on Si at different base and growth pressures. (d) The relation between the Fe valence states determined from XAS and the net magnetic moment measured by VSM. A representative error bar is shown.

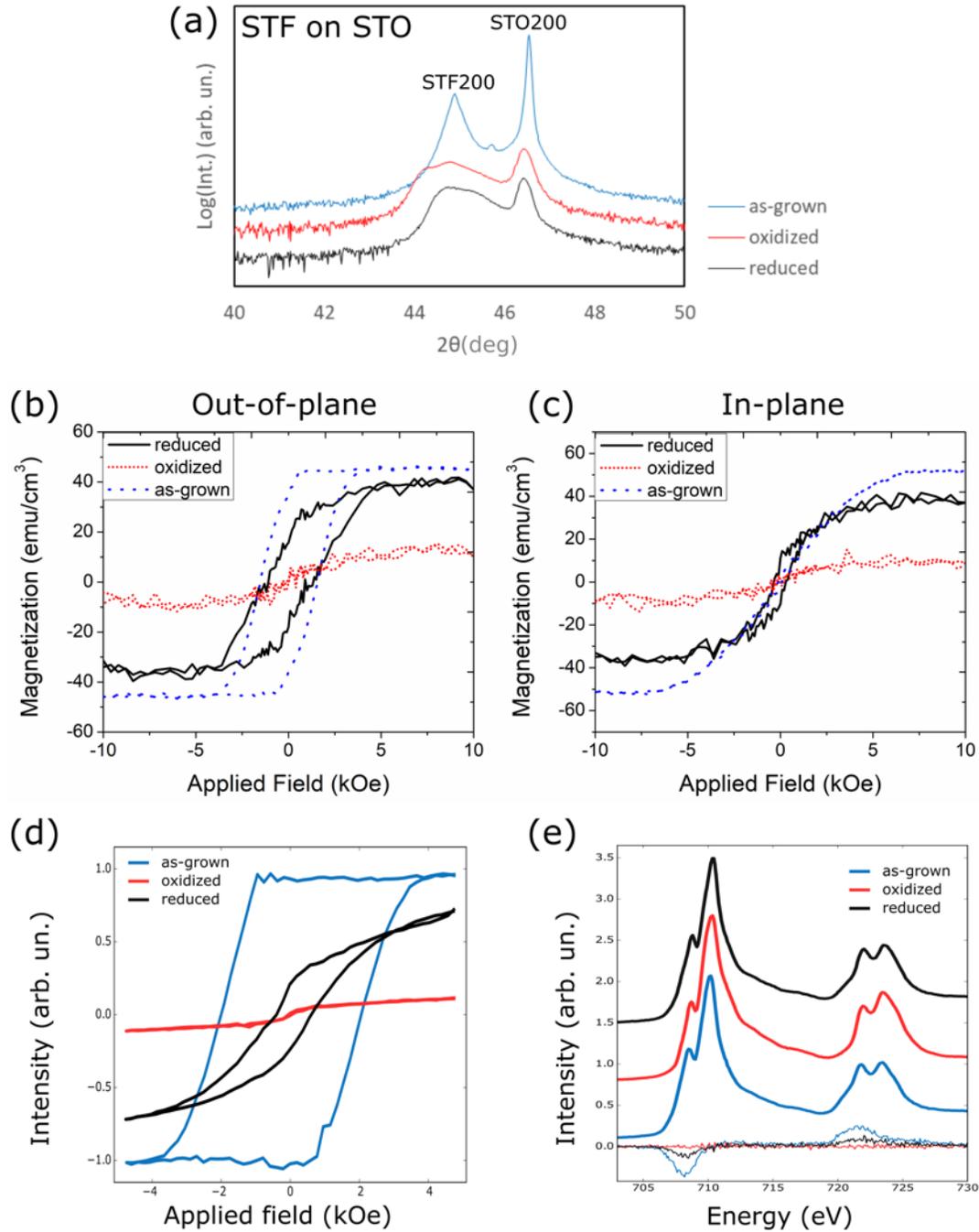

FIG. 4. (a) XRD ω-2θ scans between 40°<$2\theta$<50°, (b) out-of-plane and (c) in-plane VSM hysteresis curves for the SrTi$_{65}$Fe$_{35}$O$_{3-\delta}$ thin film subjected to various annealing treatments. (d) XMCD hysteresis curves collected on Fe signal and (e) XAS and XMCD for the same film subjected to various annealing treatments.

**Supplementary Materials for "An XMCD study of magnetism and valence state in iron-substituted strontium titanate films"**

Astera S. Tang[†], Jonathan Pelliciari[†], Qi Song, Qian Song, Shuai Ning, John W. Freeland, Riccardo Comin, Caroline A. Ross

[†]These authors contributed equally

A.S. Tang, S. Ning, Prof. C.A. Ross

Department of Materials Science and Engineering, Massachusetts Institute of Technology, Cambridge MA 02139; astera@mit.edu, caross@mit.edu

J. Pelliciari, Q. Song, Q. Song, Prof. R. Comin

Department of Physics, Massachusetts Institute of Technology, Cambridge MA 02139; jpellici@mit.edu, rcomin@mit.edu

Q. Song

State Key Laboratory of Surface Physics and Department of Physics, Fudan University, Shanghai 200433, China

J.W. Freeland

Advanced Photon Source, Argonne National Laboratory, Argonne, Illinois, 60439, USA

## Supporting Information

1. **XAS of Fe with different valence states**

    In Fig. S1, we compare the XAS spectra of different Fe states with one representative STF sample. From Fig. S1, it is clear that the Fe in this material is not present as $Fe^{4+}$, but rather as $Fe^{3+}$ with a minor fraction of $Fe^{2+}$.

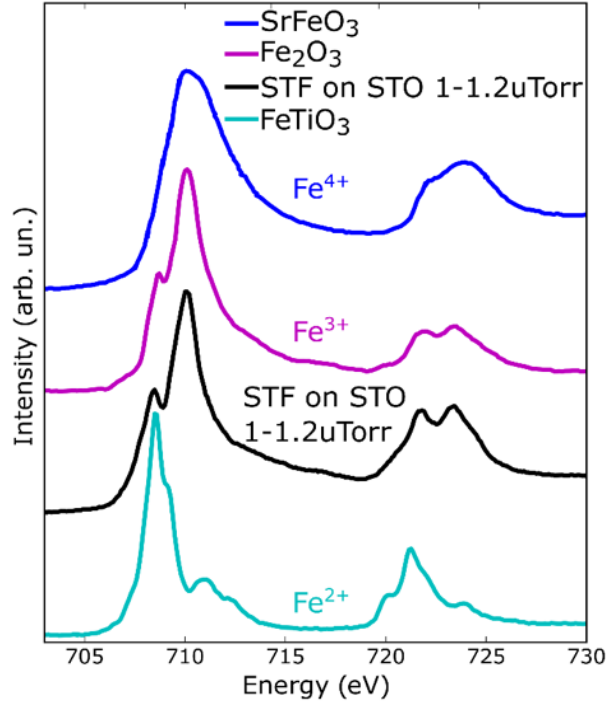

FIG. S1. Comparison of sample 1-1.2μTorr (STO) with reference compounds of SrFeO$_3$ (Fe$^{4+}$), FeTiO$_3$ (Fe$^{3+}$), and FeO (Fe$^{2+}$).

2. **Calibration procedure for the quantification of the Fe$^{2+}$/Fe$^{3+}$ fraction.**

We linearly combine the XAS spectra of Fe$_2$O$_3$ (Fe$^{3+}$) and FeTiO$_3$ (Fe$^{2+}$) using the formula

$$I_{XAS} = a \cdot I_{XAS\text{-}FeTiO3} + (1 - a) \cdot I_{XAS\text{-}Fe2O3}$$

where $a$ spans from 0 to 1. The resulting simulated XAS spectra are reported in Fig. S2, which highlights a clear change of the relative intensity of the peaks named A and B in Fig. S3.

To determine the fraction of Fe$^{3+}$ and Fe$^{2+}$ in our samples, we fit four of the simulated Fe-$L_3$ spectra in Fig. S2 (spectra with $a$ = 0.7, 0.8, 0.9, and 1) to a doublet of Lorentzian curves plus a broad Gaussian peak on the high-energy side. The Lorentzian lineshape was chosen to account for core-hole broadening of the XAS spectra and the Gaussian peak was selected to account for the broad continuum of the XAS observed at high energies. We then calculated the ratio of the areas of peak A to peak B. The calibration results are reported in Fig. S4(a,b). In Fig. S4(a), we show the ratio A/B for a wide range of $Fe^{2+}$:$Fe^{3+}$ mixtures. In Fig. S4(b), we zoomed-in on part of the same plot, focusing on the range of ratios similar to that observed in our samples, which are shown in Fig. S4(c,d). The gray dashed line marks the ratio for 1-1.2 µTorr (STO) and the "as-grown" sample, and indicates an $Fe^{3+}$:$Fe^{2+}$ ratio of 90:10. In Fig. S4(c), we compare the ratio achieved for films on Si and STO grown at different oxygen pressures. The clear decrease of the A/B ratio for samples grown with gradually higher oxygen pressure reflects the increase of relative $Fe^{3+}$ concentration. The increase of $Fe^{3+}$ is not unexpected because more oxidizing environments would naturally lead to an increase of the Fe oxidation state from 2+ to 3+.

In Fig. S4(d), we show the trend in A/B ratio for the annealed "as grown"–"oxidized"–"reduced" samples. The "as grown" sample displays both the highest A/B ratio and fraction of $Fe^{2+}$ among the three films, whereas the "oxidized" film displays the lowest. Subsequent reduction demonstrates that the source of ferromagnetism can be found in the presence of $Fe^{2+}$.

The Fe-L XAS and XMCD for STF grown on STO are shown in Fig. S5. Only the film grown at the lowest base and growth pressures within the set demonstrated magnetism.

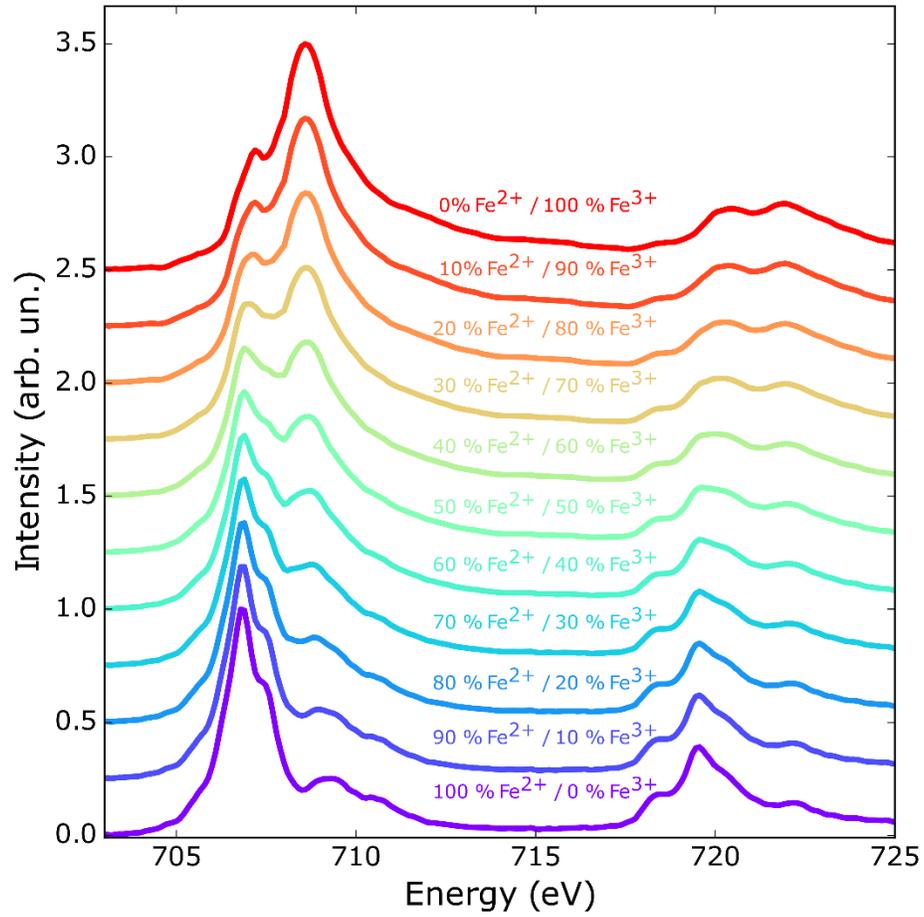

FIG. S2. Collection of XAS spectra for various compositions of $Fe^{2+}$ and $Fe^{3+}$ created by a linear combination of the XAS for $FeTiO_3$ and $Fe_2O_3$.

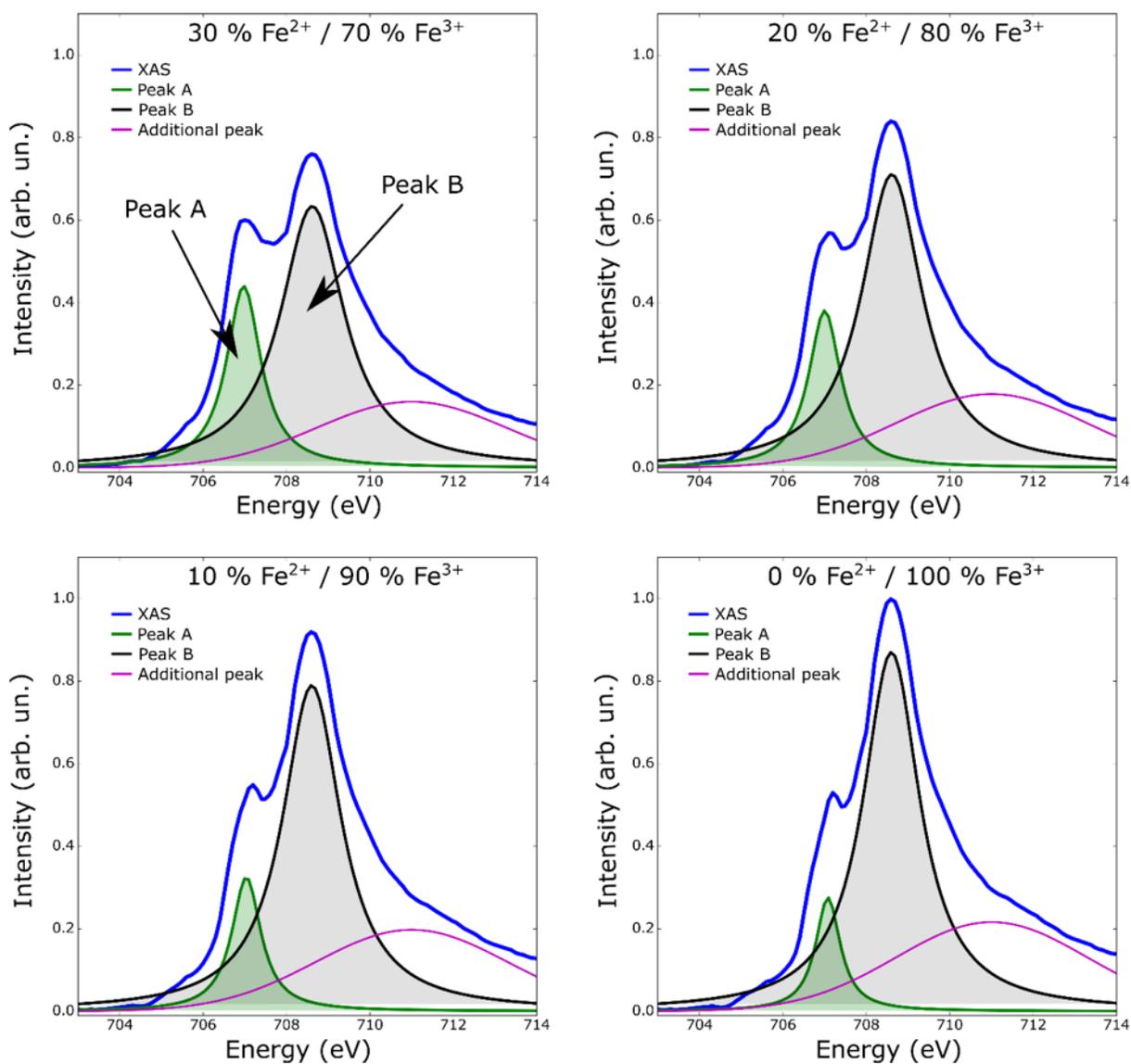

FIG. S3. Definition of peak A and Peak B and some fitting examples for different compositions of $Fe^{2+}$ and $Fe^{3+}$. Blue traces are taken from four representative simulated spectra of Fig. S2. Green and black curves represent the peak A and B respectively.

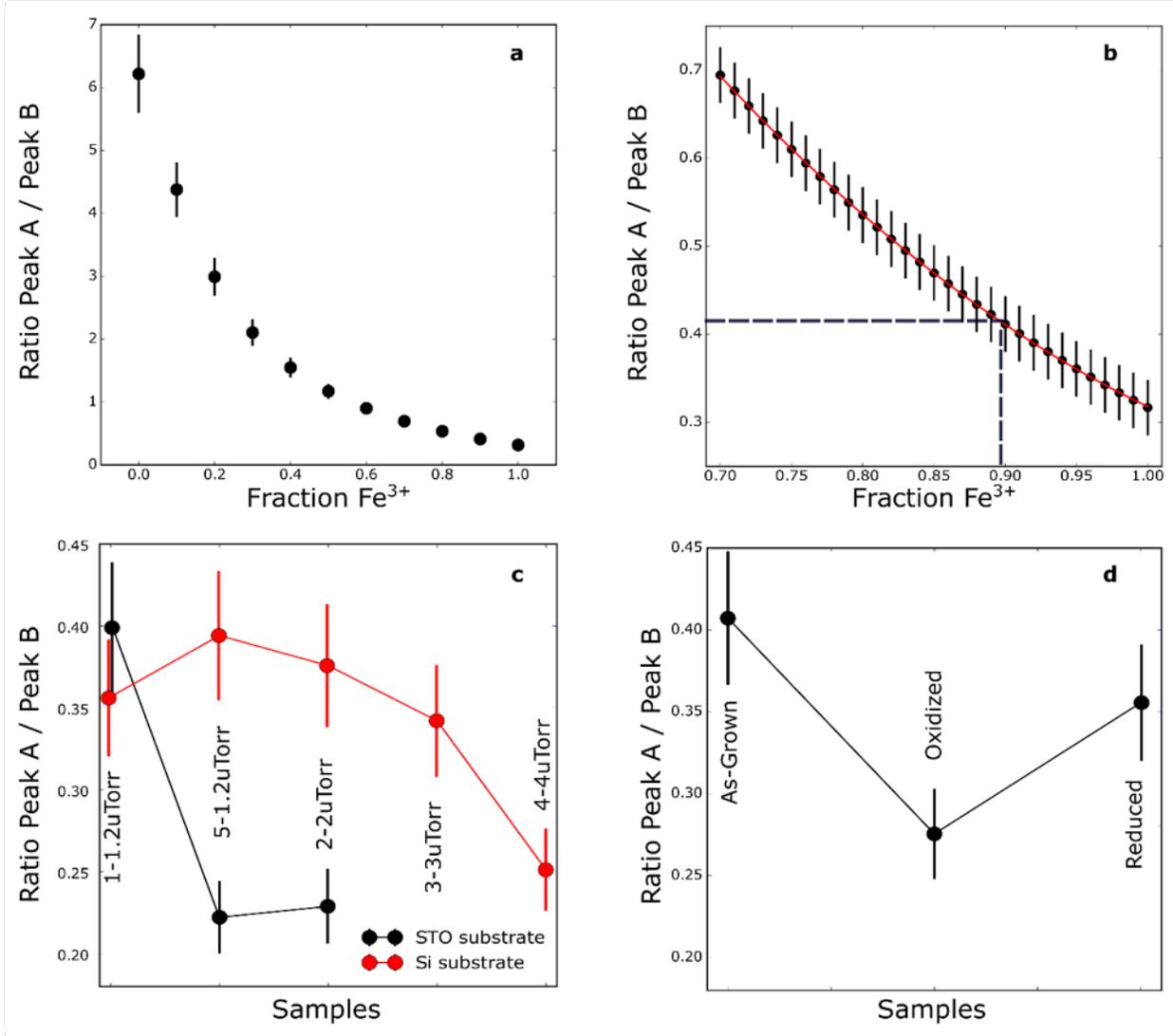

FIG. S4. Estimation of the percentage of $Fe^{2+}/Fe^{3+}$ in STF samples. (a) Wide range calibration curve used to determine the fraction of $Fe^{2+}/Fe^{3+}$. (b) Higher sampling of the calibration curve corresponding to a ratio similar to the one achieved for the samples. (c) Evolution of the $area_A/area_B$ ratio for the STF samples measured. Red points with error bars indicate samples grown on Si, black dots with error bars indicate samples grown on STO. (d) Evolution of the $area_A/area_B$ ratio for the annealed as grown-oxidized-reduced samples.

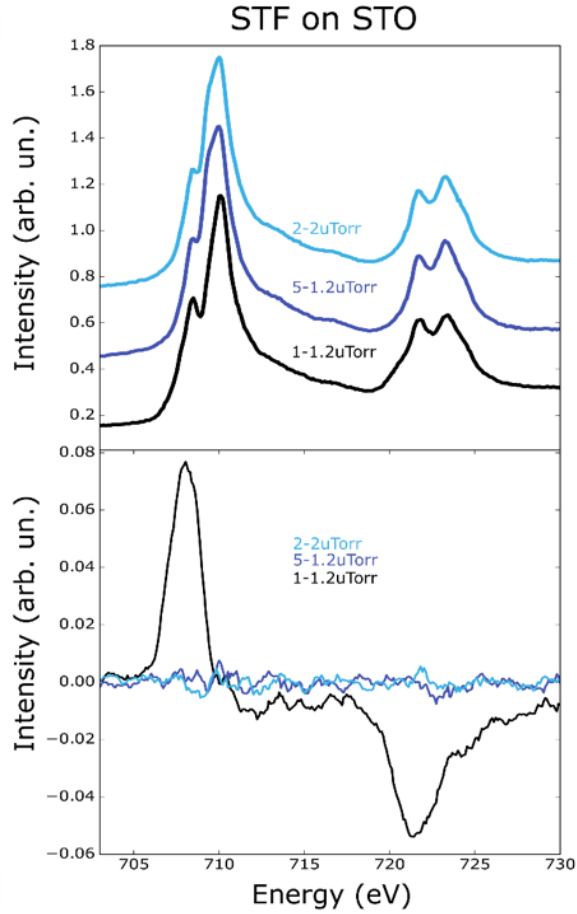

FIG. S5. Fe-L XAS and XMCD for STF grown on STO

3. **Ti-L XAS and XMCD for STF grown on Si and STO**

The XAS of Ti for both substrates is shown in Fig. S6. There is very little modification in the lineshape of the XAS spectra at different growth conditions. No detectable XMCD was observed that could implicate the Ti ions as a source of magnetism.

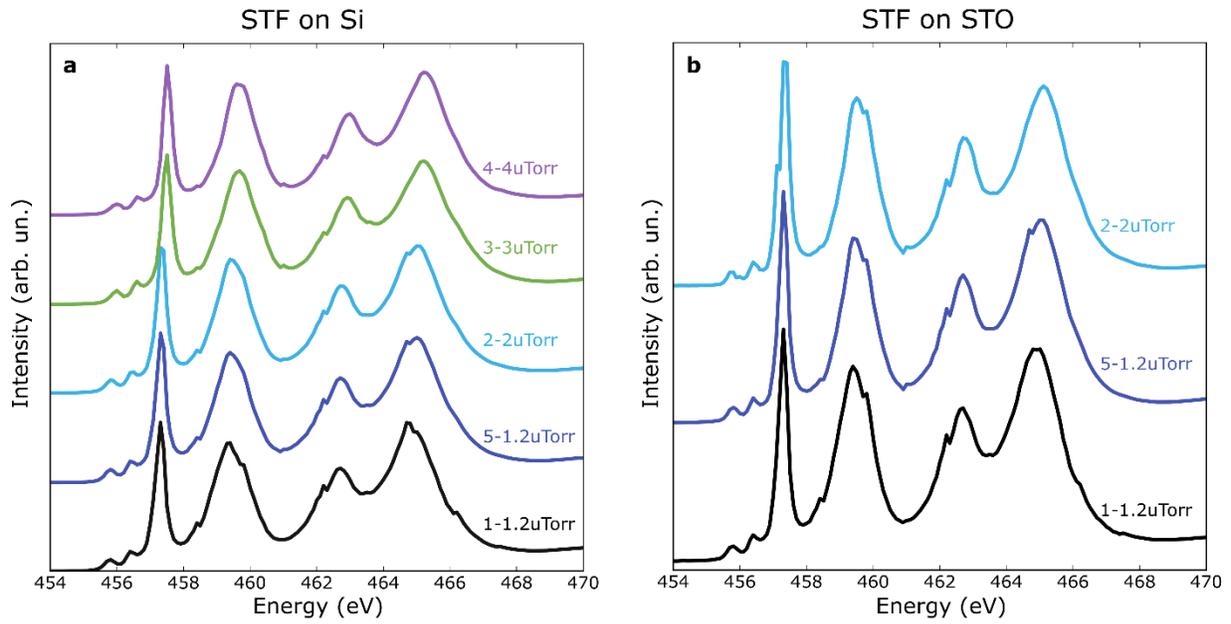

FIG. S6. Ti-L edge XAS for (a) samples grown on Si substrates and (b) samples grown on STO substrates.

4. **Qualitative comparison of XMCD signal intensity vs VSM magnetic moment**

XMCD signal intensity and VSM magnetic moment for the STF films on Si and STF films on STO are shown together in Fig. S7. There is excellent correlation between the XMCD signal and VSM magnetic moment.

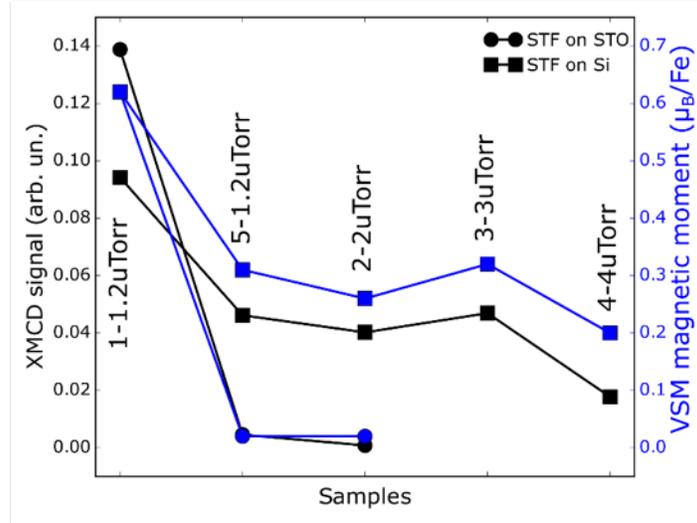

FIG. S7. XMCD signal intensity vs VSM magnetic moment for STF films on Si and STF films on STO.

### 5. TEM imaging for STF on STO

TEM images of 1-1.2μTorr (STO) are displayed in Fig. S8. Fig. S8(a,b) shows bright field (a) and dark field (b) images of the film, with the corresponding electron diffraction schematically drawn in (c). Figure S8(d) features a zoomed-in higher resolution image of the film structure. Planar defects run vertically through large sections of the film, in some cases through the entire film, and are presumably present to relieve strain.

Elemental mapping [Fig. S8(e-i)] of the same sample demonstrates that despite the preponderance of vertical defects, the distribution of the component elements is homogeneous within the film. All elements appear to be incorporated into the perovskite structure and no Fe clusters were observed, further supporting XRD data demonstrating a lack of metallic or iron oxide phases.

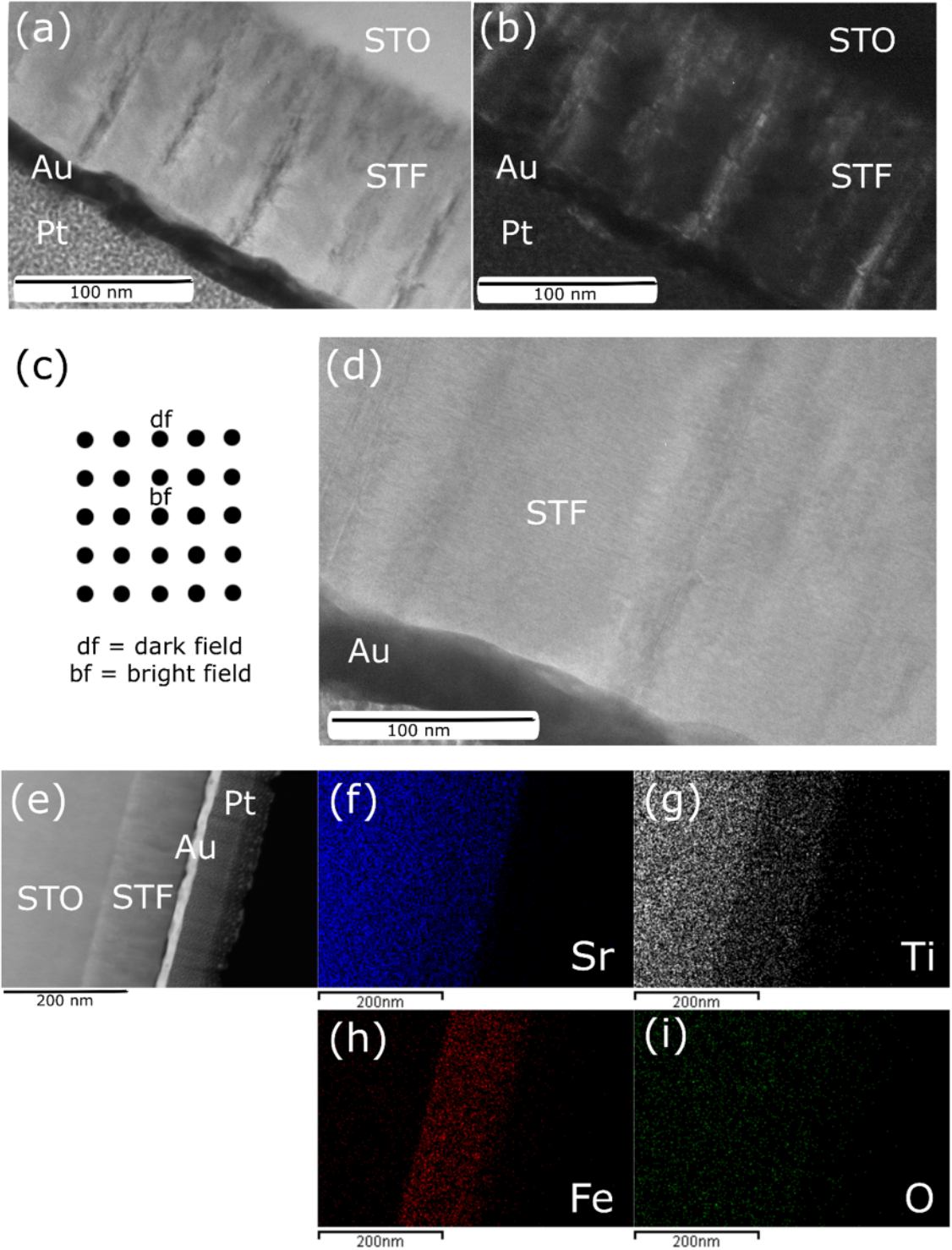

FIG. S8. (a) Bright field and (b) dark field images of the 1-1.2µTorr (STO) sample with corresponding electron diffraction schematic shown in (c). The bright field and dark field points are designated on the schematic. (d) Higher resolution image showing film structure. (e-i) Elemental mapping including (e) image of mapped area, (f) Sr mapping, (g) Ti mapping, (h) Fe mapping, and (i) O mapping.

## 6. Raman spectra for STF on Si

Raman spectra of 1-1.2µTorr (Si) and 4-4µTorr (Si) were acquired using unpolarized 532 nm light, and displayed in Fig. S9.

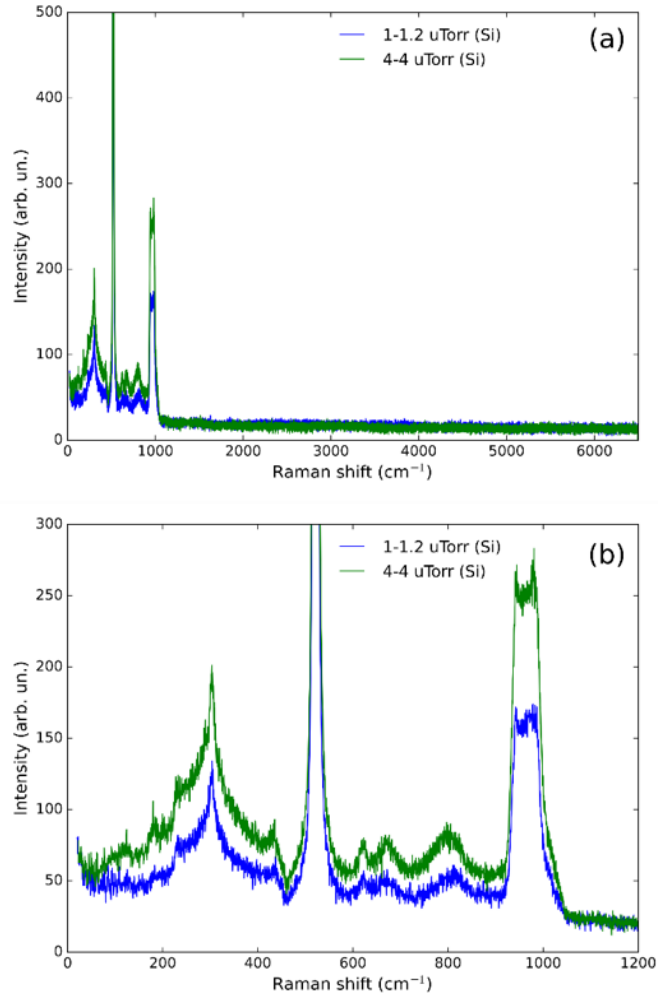

FIG. S9. (a) Raman spectrum of STF 1-1.2µTorr (Si) and 4-4µTorr (Si) in a wide range of energy transfer. (b) Expanded view of the Raman spectrum (a) in the low energy transfer region.

## 7. Reciprocal space mapping (RSM) of films grown on STO substrates

RSM scans of the (113) asymmetric peak were collected for all STF films grown on STO(100) substrates, and three are shown in Fig. S10. The color distribution follows the same scale in all scans. RSM scans show all films are strained in-plane and are epitaxial with the substrate. The RSM scans correspond well with the 2θ-ω scans. The magnetic films demonstrated a larger out-of-plane lattice parameter in both the RSM and 2θ-ω. Additionally, the annealed films showed a significantly lower film intensity than the grown films, in line with the lower peak intensity observed in the 2θ-ω scans. Due to the lower film intensity, the annealed films were scanned in grazing incidence mode, in which peak resolution is sacrificed for increased film signal. Although the film peak becomes visible with grazing incidence, the location of the peak becomes more uncertain. Nevertheless, the reduced film [Fig. S10(c)], despite having undergone two annealing treatments and experiencing the loss and partial recovery of its original magnetic signal, remains epitaxial with the substrate, albeit spread out along $q_z$. The spread in the film peak that appears upon annealing suggests a distribution of strain out-of-plane.

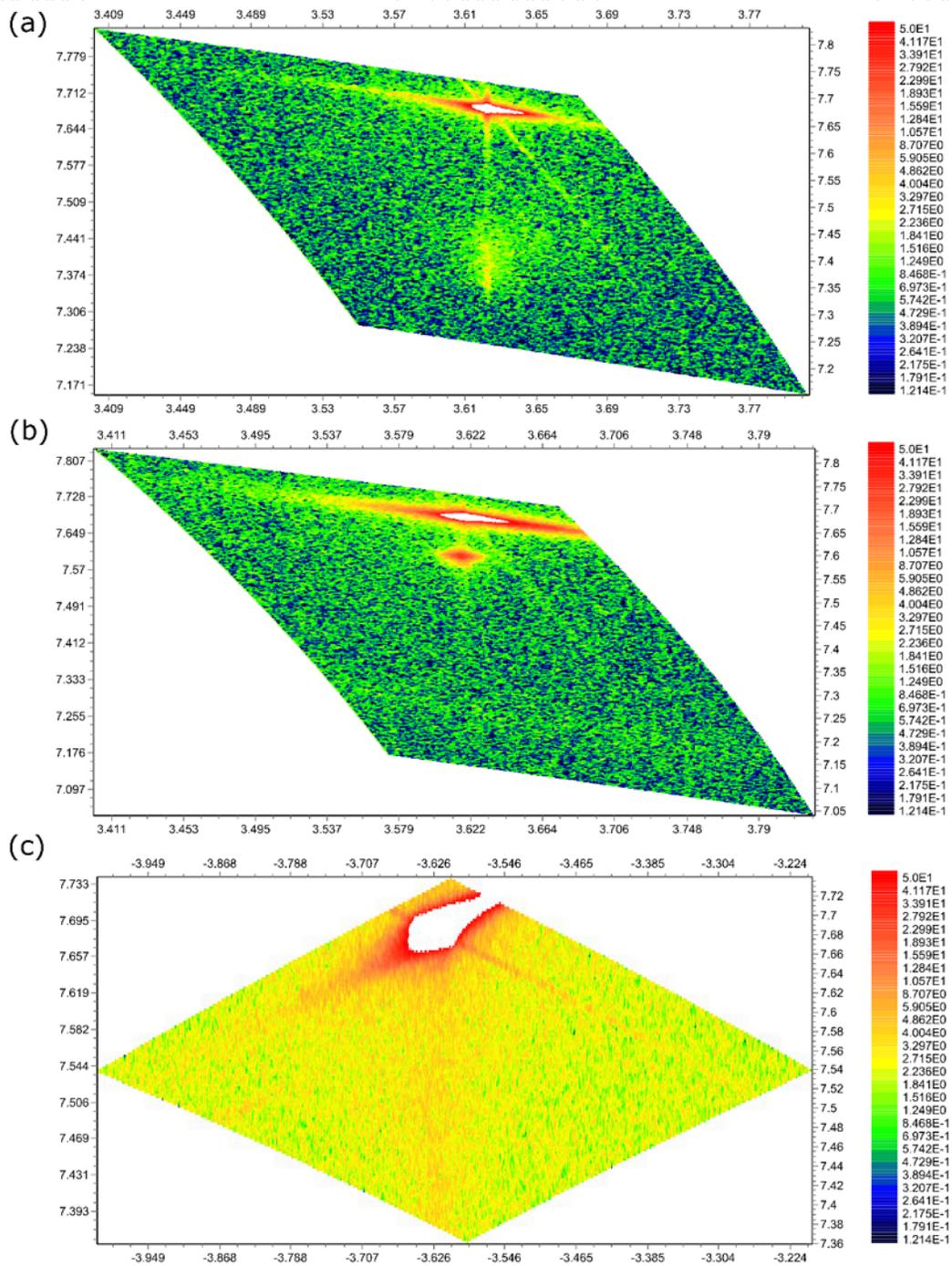

FIG S10. Reciprocal space maps for (a) 1-1.2μTorr (STO) and (b) 2-2μTorr (STO) using grazing exit setup, and (c) the reduced sample using grazing incidence setup. $q_x$ is along the x-axis, $q_z$ is along the y-axis in all maps.